\UseRawInputEncoding
\documentclass[proof]{WileyASNA-v1}

\newcommand{\arcsec}{"}

\newcommand{\kms}{km\,s$^{-1}$}
\newcommand{\ms}{m\,s$^{-1}$}

\articletype{Original Paper}%
\received{<day> <Month>, <year>}
\revised{<day> <Month>, <year>}
\accepted{<day> <Month>, <year>}

\begin{document}

\title{On the binary orbit of Henry Draper one (HD\,1)}

\author[1]{Klaus G. Strassmeier*}

\author[1]{Michael Weber*}

\authormark{Strassmeier \& Weber}

\address[1]{\orgname{Leibniz-Institute for Astrophysics Potsdam (AIP)}, \orgaddress{An der Sternwarte 16, D-14482 Potsdam, \country{Germany}}}

\corres{*\email{kstrassmeier@aip.de}, \email{mweber@aip.de}}

\abstract[Abstract]{
We present our final orbit for the late-type spectroscopic binary HD~1. Employed are 553 spectra from 13 years of observations with our robotic STELLA facility and its high-resolution echelle spectrograph SES. Its long-term radial-velocity stability is $\approx$50\,\ms . A single radial velocity of HD~1 reached a rms residual of 63\,\ms, close to the expected precision. Spectral lines of HD\,1 are rotationally broadened with $v\sin i$ of 9.1$\pm$0.1\,\kms . The overall spectrum appears single-lined and yielded an orbit with an eccentricity of 0.5056$\pm$0.0005 and a semi-amplitude of 4.44\,\kms . We constrain and refine the orbital period based on the SES data alone to 2318.70$\pm$0.32\,d, compared to 2317.8$\pm$1.1\,d when including the older data set published by DAO and Cambridge/Coravel. Owing to the higher precision of the SES data, we base the orbit calculation only on the STELLA/SES velocities in order not to degrade its solution. We redetermine astrophysical parameters for HD\,1 from spectrum synthesis and, together with the new {\it Gaia} DR-2 parallax, suggest a higher luminosity than published previously. We conclude that HD\,1 is a slightly metal-deficient K0\,III-II giant 217 times more luminous than the Sun. The secondary remains invisible at optical wavelengths. We present evidence for the existence of a third component. }


\maketitle

\footnotetext{\textbf{Abbreviations:} STELLA Stellar Activity, SES STELLA Echelle Spectrograph}

% ==============================================================================================================================

\section{Introduction}\label{S1}

Attracted by the star's outstanding Henry Draper (HD) number and the fact that almost nothing substantial was known about it, we decided to monitor the target on our STELLA robot in Tenerife starting in the year 2007. It soon turned out that HD~1 is a long-period spectroscopic binary and we wrapped up our initial collection of STELLA data for a first paper including a preliminary orbit (Strassmeier et al.~2010, hereafter Paper~I). What we did not know at that time was that the target had been observed by R.~McClure at the Dominion Astrophysical Observatory (DAO) in Canada from 1980 until 1994, and was continued thereafter by R. Griffin in Cambridge, U.K., from 2002 until 2008. At the time we had submitted our Paper~I, Griffin \& McClure (2009) had independently derived an orbit from their data and published it in The Observatory. No arXive or other info across the continent was available then, so we missed it. We apologize deeply.

Almost needless to say, although relaxing, is that the two orbits agreed very well. The high eccentricity and even the periastron longitude, usually a not well defined element, agreed to within its errors. Only marginal differences were seen for the systemic velocity and the semi-amplitude, probably owing to zero-point differences between the three telescope-spectrograph systems. While the DAO/Cambridge data set had a much longer baseline in time than ours (28 years viz. 3 years), our data were more precise by more than a factor of four as judged from the root-mean-square (rms) residual of a single radial velocity (RV) of unit weight with respect to the orbital solution (73\,\ms\ vs. 310\,\ms ). Consequently, and despite of the large difference in the baseline, even the orbital periods agreed to within the respective errors, but was uncomfortably uncertain in our case ($\pm$69~d) compared to the Griffin \& McClure (2009) orbit ($\pm$2.7~d).

In this paper, we present our final orbit of HD~1. We add 10 more years of STELLA monitoring of the target with the same telescope-spectrograph combination as in Paper~I, and now have 553 echelle spectra and high-precision radial velocities available.

%---------------------------------- F1
\begin{figure*}
\center
\includegraphics[angle=0,width=\textwidth,clip]{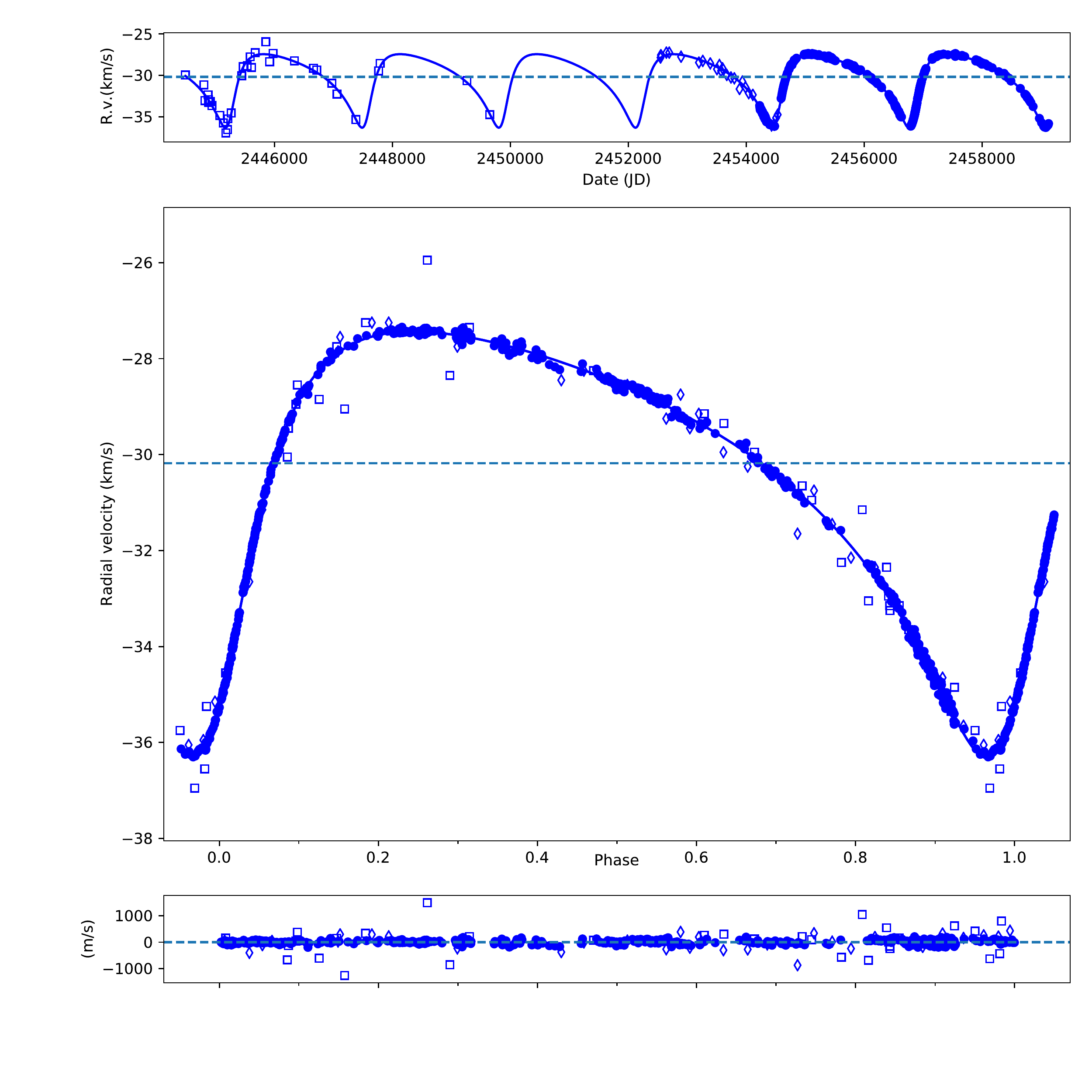}
\caption{Radial velocities and orbit for HD\,1. The dashed horizontal line is the systemic velocity. The filled dots are the STELLA/SES velocities, the open squares are the DAO and the open diamonds the Cambridge/Coravel velocities. A total of 553 SES, 34 DAO and 30 Cambridge/Coravel velocities were employed. The full line is the orbital solution from the SES data alone. }\label{F1}
\end{figure*}

\section{Observations and data reduction}

All observations were made with the two 1.2\,m STELLA telescopes on Tenerife in the Canary islands and the fiber-fed echelle spectrograph SES (Strassmeier et al. 2004, Weber et al. 2012). Up to 2012 SES was fed from one of the two Nasmyth foci of STELLA-I. In early 2012 the fiber was moved to the prime focus of STELLA-II. The spectra cover the full optical wavelength range from 390--880\,nm with an average resolving power of $R$=60\,000 at a sampling of two pixels per resolution element. SES employs an e2v 2k$\times$2k CCD detector. The slightly higher spectral resolution compared to the earlier data in Paper~I (55,000) was made possible with a 2-slice image slicer and a 50-$\mu$m fiber resulting in a projected sky aperture of 3.7\arcsec . At a wavelength of 600\,nm this corresponds to an effective resolution of 5.5\,\kms\ or 110\,m\AA . The average dispersion is 45~m\AA/pixel. Further details of the performance of the system can be found in previous applications to comparable stars, for example in Weber \& Strassmeier (2011) or Strassmeier et al. (2020).

For HD\,1, we adopted an integration time of 1200\,s. The signal-to-noise ratios (S/N) achieved ranged between 70:1 and 300:1 per pixel, depending on weather conditions. A full SES example spectrum was shown in Paper~I. The first spectrum of HD~1 was taken on May~6, 2007, the last one on Oct.~5, 2020, thus covering more than 13 years. Our initial STELLA observations for the preliminary orbit in Paper~I were from 2007 to early 2010. We thus add 10 more years of data in the present paper. The total number of CCD frames obtained was 602, of which 29 ($\approx$5\%) had to be discarded because of too low a S/N due to thick clouds or CCD cooling interruptions. From the remaining 573 radial velocity measurements, 20 were discarded by the 3-sigma-clipping during calculation of the orbit, leaving 553 for the final orbit calculation. All these 20 discarded data points had low S/N, typical of spectra taken through thin clouds.

SES spectra are reduced automatically using the IRAF-based STELLA data-reduction pipeline (Weber et al. 2008). The images were corrected for bad pixels and cosmic-ray impacts. Bias levels were removed by subtracting the average overscan from each image followed by the subtraction of the mean of the master bias frame (from which the overscan had already been subtracted). The target spectra were flattened by dividing by a nightly master flat which has been normalized to unity. The nightly master flat itself is constructed from around 50 individual flats observed during dusk, dawn, and around midnight. After removal of scattered light, the one-dimensional spectra were extracted using an optimal-extraction algorithm. The blaze function was then removed from the target spectra, followed by a wavelength calibration using consecutively recorded Th-Ar spectra. Finally, the extracted spectral orders were continuum normalized by dividing with a flux-normalized synthetic spectrum of comparable spectral classification as the target.

\section{Orbit determination}\label{S}

Radial velocities from the STELLA spectra were determined from an order-by-order cross correlation with a synthetic template spectrum calculated using MARCS atmospheres (Gustafsson et al.~2008), the Turbospectrum code (Plez 2012), and a line list taken from VALD (Ryabchikova et al.~2015) with parameters that matched HD~1. Sixty out of the 82 echelle orders were used for the cross correlation, the individual cross-correlations weighted according to the spectral region and averaged, and the radial velocity measured by a fit to this average function. Its rms defines our internal RV error ($\approx$8~\ms ). The external rms values are naturally significantly larger and strongly depend on the star, its line broadening and blending. All RVs in this paper are barycentric and are corrected for Earth rotation. A complete list of our RV data is available at CDS\footnote{via anonymous ftp to cdsarc.u-strasbg.fr (130.79.128.5) or via http://cdsweb.u-strasbg.fr/cgi-bin/}.

As in Paper~I, we solved for the elements of a spectroscopic binary using the general least-squares fitting algorithm MPFIT (Markwardt 2009). For solutions with non-zero eccentricity, as in this paper, we used the prescription from Danby \& Burkardt (1983) for the calculation of the eccentric anomaly. The data and the orbital fit are shown in Fig.~\ref{F1}. The orbital elements are listed in Table~\ref{T1}.

%------------------------------ Table 1: Orbital elements
\begin{table}[!tbh]
\begin{flushleft}
\caption{Orbital elements of HD~1.}\label{T1}
\begin{tabular}{lll}
\hline\noalign{\smallskip}
Orbital element          & SES alone                         & All data\\
                         & (=final orbit)                    &     \\
\noalign{\smallskip}\hline\noalign{\smallskip}
$P$ (days)               & $ 2318.70 \pm 0.32 $              & $ 2317.8 \pm 1.1 $ \\
$T_0$ (HJD 245+)         & $ 6,837.88 \pm 0.44 $             & $ 6,837.6 \pm 1.2 $ \\
$\gamma$ (\kms)          & $ -30.145 \pm 0.003 $             & $ -30.177 \pm 0.007 $ \\
$K_{1}$ (\kms)           & $ 4.446 \pm 0.004 $               & $ 4.435 \pm 0.013 $ \\
$e$                      & $ 0.5056 \pm 0.0005 $             & $ 0.503 \pm 0.002 $\\
$\omega$ (deg)           & $ 220.97 \pm 0.15 $               & $ 221.0 \pm 0.4 $ \\
$a_1\sin i$\,($10^6$\,km)& $ 122.3 \pm 0.13 $                & $ 122.2 \pm 0.4 $ \\
$f(M)$ (M$_\odot$)       &   0.01359$\pm$0.00004             & $ 0.0136 \pm 0.0001 $ \\
$rms$ (\ms)              & 63                                & 155 \\
$N$                      & 553                               & 622 \\
\noalign{\smallskip}\hline
\end{tabular}
\end{flushleft}
\end{table}

For the present orbit determination, we took advantage of the Griffin \& McClure (2009) data set. It soon turned out, though, that their data precision per observation was actually degrading the STELLA solution. Griffin \& McClure (2009) had to bring their two data sets first into systematic agreement by shifting the DAO velocities by +1.2\,\kms\ with respect to the Cambridge/Coravel data. Additionally, and to account for the diverging data precision, they assigned a weight of 0.4 to the DAO data and unity to the Cambridge/Coravel data.

In the SES data, we found a linear trend of $-0.011\pm 0.002$~\ms\ per day in the residuals of the orbital solution, which cause an $\approx$50\,\ms\ offset between the first and the last RV point in the STELLA time coverage. The origin of the trend remains unknown but could be due to an unseen third component with a very long orbital period but in principle also due to a hitherto unknown time-dependent instrumental drift. Because our RV standard stars do not show such a drift, we are inclined to believe that we see the indirect signature of a third component. The interpolated RV values from the fit are used as a correction to all SES RV points for the final orbital solution.

\begin{figure}
\center
\includegraphics[angle=0,width=90mm,clip]{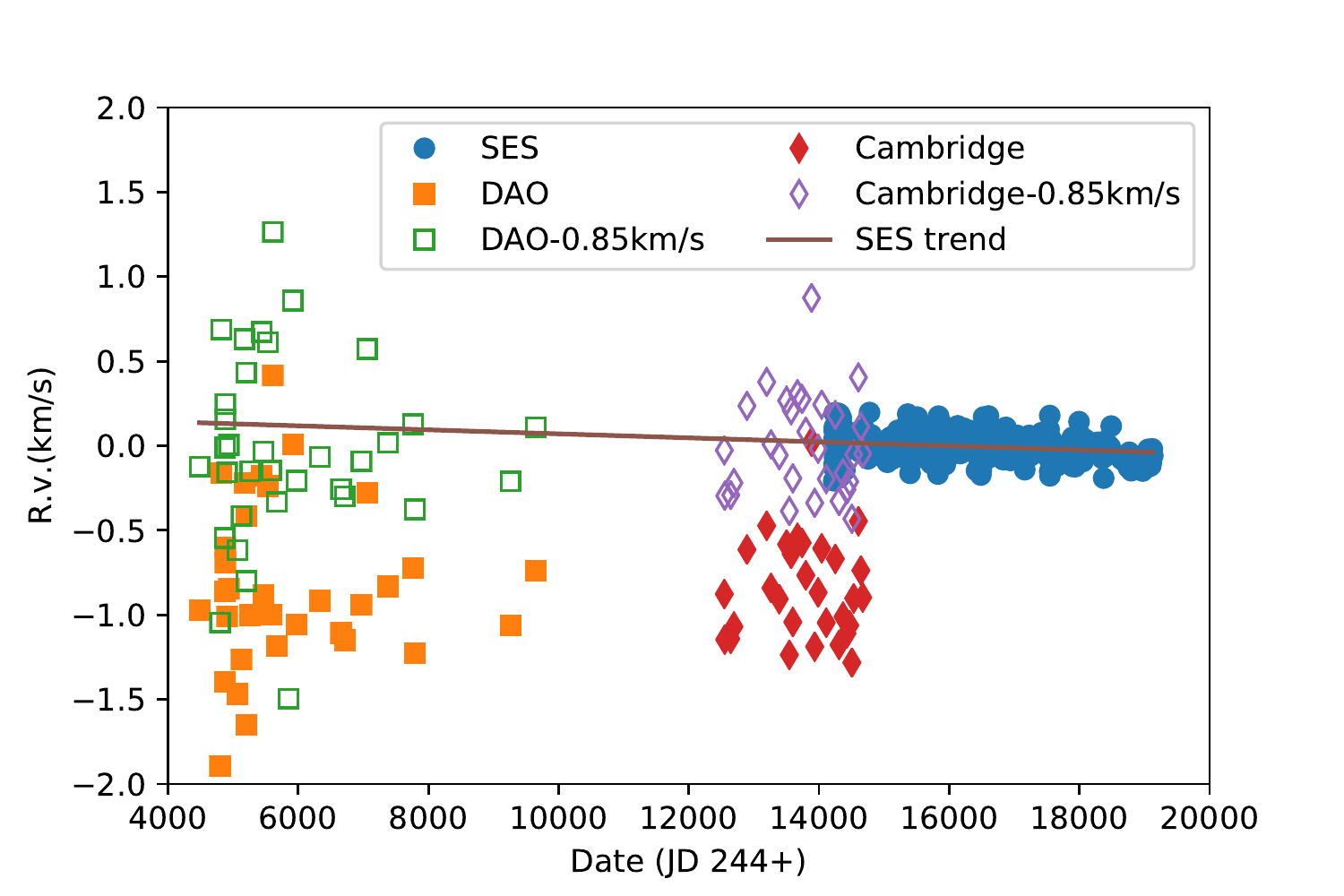}
\caption{RV residuals of the three data sets versus JD with respect to the final SES orbit. Shown are the RVs before and after the shift applied. Symbols are as in Fig.~\ref{F1}. The original unshifted data are represented by the same symbol type but are filled-in. The line is a fit to the linear trend of the SES data.}\label{F2}
\end{figure}

We kept the relative shift of the DAO velocities with respect to the Cambridge/Coravel velocities, as suggested by Griffin \& McClure (2009), but shifted both data sets by $-0.85$\,\kms\ in order to match the SES velocities. The new shift was determined from a minimization of the Cambridge/Coravel velocity residuals with respect to the SES orbit when they overlapped in time with STELLA observations (HJD 2,454,250.59--671.61). The lower-precision DAO velocities do not align perfectly with but also do not contradict the trend the SES and Cambridge/Coravel velocities suggest but are too noisy for a conclusive verification. We note that the SES velocities are on the IAU system defined by Udry et al. (1999). Figure~\ref{F2} shows the residuals of all three data sets and their individual shifts applied. The line is the linear fit to the SES trend mentioned above.

The final elements are derived from SES data alone and reach a rms residual of 63\,\ms\ for a data point of unit weight. The lowered precision and the uncertain zero points for the Griffin \& McClure (2009) data set forced us to decide not using the DAO/Cambridge data for the final orbital solution but only for the initial period determination. Nevertheless, we give in the second column of Table~\ref{T1} the orbital elements that we derive if all data are employed with equal weight and the shifts as shown in Fig.~\ref{F2}. The overall rms of its solution is 155\,\ms . The two orbital solutions agree within their errors, but since the precision of even the period determination is so much better in the SES solution, we decided to also use that period despite the shorter time coverage. At this point we also note that the STELLA/SES data in the present paper were shifted with a slightly different value compared to Paper~I in order to match the IAU/Coravel zero point defined by Udry et al. (1999). The new STELLA RV zero point was obtained by Strassmeier et al. (2012) as +0.503\,\kms\ from long-term observations of 22 RV standard stars. This value had been added to all RVs of HD\,1 sent to CDS.

%-------------------------   Table 2:  Astrophysical values of HD 1
%par 0: -0.155000 +/- 0.040000 (std 0.007481, std of err is 0.007035)
%par 1: 1.813781 +/- 0.008599 (std 0.023397, std of err is 0.001650)
%par 2: 9.107152 +/- 0.014629 (std 0.142266, std of err is 0.002496)
%par 3: 4881.827000 +/- 35.848000 (std 5.922214, std of err is 6.218753)
%par 4: 2.184000 +/- 0.133000 (std 0.025729, std of err is 0.023191)
\begin{table}[!tb]
\begin{flushleft}
\caption{Summary of astrophysical data of HD~1. \label{T2}}
 \begin{tabular}{ll}
  \hline\noalign{\smallskip}
  Parameter                & Value    \\
  \noalign{\smallskip}\hline\noalign{\smallskip}
  $V$-magnitude (average)  & 7.37 mag \\
  Spectral classification  & K0\,III-II  \\
  {\it Gaia} DR-2 distance & $365 \pm 5$~pc \\
  $M_V$                    & $-0.73 \pm 0.03$ mag\\
  $A_V$                    & $0.29$ mag\\
  $T_{\rm eff}$            & $4882 \pm 36$ K   \\
  $\log g$                 & $2.18 \pm 0.13$ cm\,s$^{-2}$\\
  $v\sin i$                & $9.11 \pm 0.02$ km\,s$^{-1}$   \\
  $[$M/H$]$                & $-0.16 \pm 0.04$ \\
  $\xi$                    & $1.81 \pm 0.01$ \kms \\
  Radius                   & $20.7 \pm 0.3$ R$_\odot$  \\
  Luminosity               & $217 \pm 6$ L$_\odot$ \\
  Mass                     & $3.5 \pm 0.1$ M$_\odot$  \\
  Age                      & $\approx$300 Myr  \\
  \noalign{\smallskip}\hline
 \end{tabular}
\end{flushleft}
{{\bf Note:} Errors for the spectrum synthesis values ($T_{\rm eff}$, $\log g$, $[$M/H$]$, $v\sin i$, $\xi_t$) are internal errors. See text.}
\end{table}

%---------------------------------- F2

\section{Discussion}\label{D}

Precise astrophysical parameters of giant stars are uncertain because there are only a few such stars that can be studied in spectroscopic binary systems, and even fewer are also eclipsing binaries. Roche lobe overflow and mass exchange usually set a limit to the stellar radius in a close binary. For HD~1, we selected 26 spectral regions from the STELLA/SES spectra covering the range 543--754\,nm and used them to redetermine the stellar effective temperature $T_{\rm eff}$, the gravity $\log g$, the metallicity $[$M/H$]$, the rotational broadening $v\sin i$, and the microturbulence $\xi_t$. We applied the program ParSES (PARameters from SES) which is based on the synthetic spectrum fitting procedure laid out by Allende-Prieto et al. (2006). Model spectra were again calculated using MARCS, Turbospectrum, and an initial VALD line-list. Synthetic spectra were pre-tabulated for a large parameter and wavelength range. This grid was then used to compare with the selected spectral regions of the individual HD~1 spectra. For the final fit, we adopted the {\it Gaia}-ESO ``clean'' line-list (Jofre et al.~2014) with windows of varying widths around the line cores of between $\pm$0.05 to $\pm$0.25\,\AA . The number of free parameters was five ($T_{\rm eff}$, $\log g$, $[$M/H$]$, $v\sin i$, $\xi_t$). Table~\ref{T2} lists the results. These values supersede the values given in Paper~I. Internal errors are obtained from the rms of the fits to the various selected wavelength regions. The more relevant external errors are estimated from comparisons with benchmark stars (see Strassmeier et al.~2018). For the present spectra the external error for $T_{\rm eff}$ is 70~K, for $\log g$ 0.1~dex, and for $[$M/H$]$ typically below 0.1\,dex.

%---------------------------------- F3
\begin{figure}
\center
\includegraphics[angle=0,width=90mm,clip]{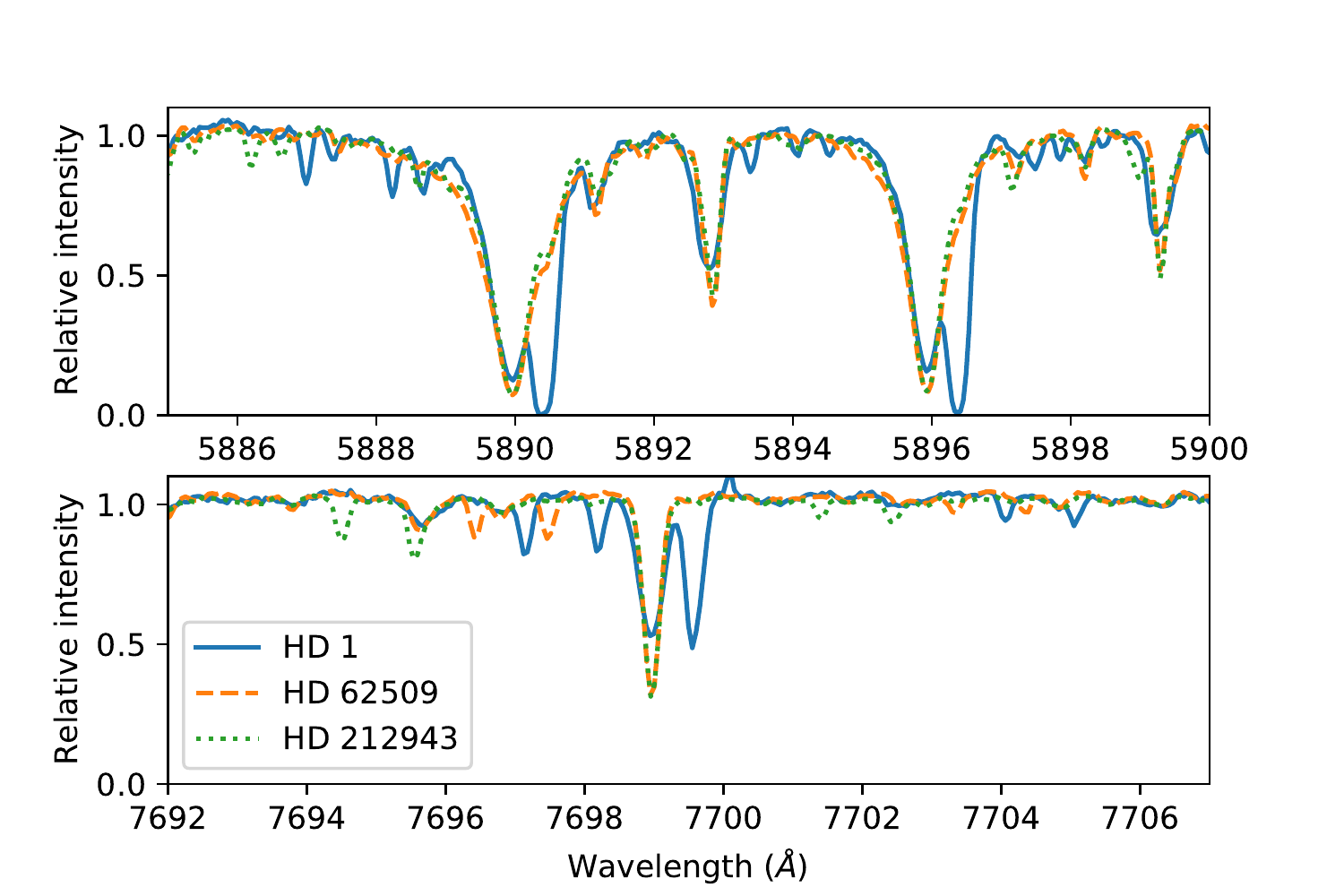}
\caption{Interstellar absorption lines in the spectrum of HD\,1. Top: the sodium Na\,{\sc i} D$_1$ and D$_2$ lines around 5890\,\AA . Bottom: the potassium K\,{\sc i} line at 7699\,\AA . The respective ISM absorptions appear redshifted with respect to the photospheric lines and a few per cents deeper. Overplotted are two nearby reference stars of similar spectral classification ($\beta$\,Gem=HD\,62509, and 35\,Peg=HD\,212943). All spectra are shifted to the rest wavelength.}\label{F3}
\end{figure}

The second ESA/{\it Gaia} data release (DR-2; Gaia collaboration 2018) gave a parallax of 2.7368$\pm$0.0371\,mas for HD\,1, which puts the star at a distance of 365$\pm$5~pc.  The previous best parallax, the revised {\it Hipparcos} parallax of 3.20$\pm$0.48~mas (van Leeuwen 2007), placed HD~1 at a distance of 312$^{+55}_{-41}$~pc, that is closer by 53~pc and only 30~pc closer than the originally published {\it Hipparcos} parallax of 2.53$\pm$0.69~mas by ESA (1997). The {\it Gaia} parallax has by far the superior precision though. However, the new distance increases even more the discrepancy of the absolute brightness from the parallax and the one expected from a nominal K0\,III giant inferred from the spectral synthesis in Paper~I. The revised absolute brightness of HD\,1 is now $M_{\rm V}$ =$-0.73\pm0.03$~mag. We applied the same mean extinction value from Henry et al. (2000) as we had used in Paper~I ($A_{\rm V}$=0.8 mag\,kpc$^{-1}$). With $T_{\rm eff}=4882$~K from the spectrum synthesis in this paper the bolometric magnitude $M_{\rm bol}$ is $-1.09$\,mag and, with a solar bolometric magnitude of +4.75\,mag, the luminosity of HD\,1 is 217$\pm$6~L$_\odot$. The Stefan-Boltzmann law then constrains the stellar radius to 20.7$\pm$0.3~R$_\odot$. These values are better matched by a luminosity class of III-II instead of III, as suggested previously from the spectrum synthesis. It is in agreement with various luminosity sensitive spectral line ratios (Strassmeier \& Fekel 1990) and also better explains the presence of strong interstellar absorption seen in the Na\,{\sc i} D and the K\,{\sc i} doublets. Figure~\ref{F3} shows this in comparison with two nearby giants of similar spectral classification. Because HD~1 falls on the ascent of the red giant branch (RGB), its mass inferred from a comparison with evolutionary tracks is uncertain, in particular because the star appears to be significantly metal deficient with $[$M/H$]$=$-0.16$. Following Paper~I, we estimate a mass of $\approx$3.5~M$_\odot$ at an age of $\approx$300\,Myr, if on the first ascent of the RGB. Table~\ref{T2} summarizes the revised astrophysical parameters of HD~1. With the new {\it Gaia} parallax, HD~1 is slightly more massive and more luminous than a nominal K0 class-III giant. It may even be a triple system.

\subsection{Acknowledgements}

STELLA was funded by the Science and Culture Ministry of the German State of Brandenburg (MWFK) and the German Federal Ministry for Education and Research (BMBF), and is operated by AIP jointly with IAC. We thank the Centre de Donn\'ees astronomiques de Strasbourg (CDS) for their services. Also appreciated are the many very helpful comments from an anonymous referee which resulted in an improved paper.

%{Bibliography}
\bibliographystyle{Wiley-ASNA}

\end{document}